\documentclass[pdflatex,sn-mathphys-num]{sn-jnl}
\usepackage{graphicx}%
\usepackage{multirow}%
\usepackage{amsmath,amssymb,amsfonts}%
\usepackage{amsthm}%
\usepackage{mathrsfs}%
\usepackage[title]{appendix}%
\usepackage{xcolor}%
\usepackage{textcomp}%
\usepackage{manyfoot}%
\usepackage{booktabs}%
\usepackage{algorithm}%
\usepackage{algorithmicx}%
\usepackage{algpseudocode}%
\usepackage{listings}%
\usepackage[figurename=Figure]{caption}
\usepackage{caption}
\usepackage[separate-uncertainty=true]{siunitx}
\DeclareSIUnit\eVolt{eV}
\usepackage[version=3]{mhchem}

\theoremstyle{thmstyleone}%

\theoremstyle{thmstyletwo}%

\theoremstyle{thmstylethree}%

\begin{document}

\title{Cavity electrodynamics of van der Waals heterostructures}

\author[1]{\fnm{Gunda} \sur{Kipp}}
\equalcont{These authors contributed equally to this work.}
\author[1]{\fnm{Hope M} \sur{Bretscher}}
\equalcont{These authors contributed equally to this work.}
\author[1,2]{\fnm{Benedikt} \sur{Schulte}}
\author[1]{\fnm{Dorothee} \sur{Herrmann}}
\author[1,2]{\fnm{Kateryna} \sur{Kusyak}}
\author[1,2]{\fnm{Matthew W} \sur{Day}}
\author[1]{\fnm{Sivasruthi} \sur{Kesavan}}
\author[1]{\fnm{Toru} \sur{Matsuyama}}
\author[1]{\fnm{Xinyu} \sur{Li}}
\author[1]{\fnm{Sara Maria} \sur{Langner}}
\author[1]{\fnm{Jesse} \sur{Hagelstein}}
\author[2]{\fnm{Felix} \sur{Sturm}}
\author[1]{\fnm{Alexander M} \sur{Potts}}
\author[1,3]{\fnm{Christian J} \sur{Eckhardt}}
\author[2]{\fnm{Yunfei} \sur{Huang}}
\author[4]{\fnm{Kenji} \sur{Watanabe}}
\author[5]{\fnm{Takashi} \sur{Taniguchi}}
\author[1,6,7]{\fnm{Angel} \sur{Rubio}}
\author[1,3]{\fnm{Dante M} \sur{Kennes}}
\author[8,1]{\fnm{Michael A} \sur{Sentef}}
\author[9,10]{\fnm{Emmanuel} \sur{Baudin}}
\author[1]{\fnm{Guido} \sur{Meier}}
\author[1]{\fnm{Marios} \sur{H Michael}}
\author*[1,2]{\fnm{James W} \sur{McIver}}
 \affil*[]{Corresponding author}

\affil[1]{\orgname{Max Planck Institute for the Structure and Dynamics of Matter},\orgaddress{\city{ Hamburg}, \country{Germany}}}
\affil[2]{\orgdiv{Department of Physics}, \orgname{Columbia University}, \orgaddress{\city{New York, NY}, \country{USA}}}
\affil[3]{\orgdiv{Institut für Theorie der Statistischen Physik}, \orgname{RWTH Aachen University and JARA-Fundamentals of Future Information Technology}, \orgaddress{\city{Aachen},\country{ Germany}}}

\affil[4]{\orgdiv{Research Center for Electronic and Optical Materials}, \orgname{National Institute for Materials Science}, \orgaddress{\city{Tsukuba},\country{ Japan}}}
\affil[5]{\orgdiv{Research Center for Materials Nanoarchitectonics}, \orgname{National Institute for Materials Science}, \orgaddress{\city{Tsukuba},\country{ Japan}}}
\affil[6]{\orgdiv{Center for Computational Quantum Physics}, \orgname{Simons Foundation Flatiron Institute}, \orgaddress{\city{New York, NY},\country{ USA}}}
\affil[7]{\orgdiv{CNano-BioSpectroscopy Group, Departamento de Fisica de Materiales}, \orgname{Universidad del País Vasco}, \orgaddress{\city{San Sebastián},\country{ Spain}}}

\affil[8]{\orgdiv{Institute for Theoretical Physics and Bremen Center for Computational Materials Science}, \orgname{University of Bremen}, \orgaddress{\city{Bremen},\country{ Germany}}}

\affil[9]{\orgdiv{Laboratoire de Physique de l'Ecole Normale Sup\'erieure}, \orgname{Universit\'e PSL, CNRS, Sorbonne Universit\'e, Universit\'e Paris-Cit\'eo}, \orgaddress{\city{Paris},\country{ France}}}
\affil[10]{\orgdiv{Institut Universitaire de France}, \orgaddress{\country{France}}}

\abstract{

Van der Waals (vdW) heterostructures host many-body quantum phenomena that can be tuned \textit{in situ} using electrostatic gates. These gates are often microstructured graphite flakes that naturally form plasmonic cavities, confining light in discrete standing waves of current density due to their finite size. Their resonances typically lie in the GHz~-~THz range, corresponding to the same \SI{}{\micro\eVolt}~-~meV energy scale characteristic of many quantum effects in the materials they electrically control. This raises the possibility that built-in cavity modes could be relevant for shaping the low-energy physics of vdW heterostructures. However, capturing this light-matter interaction remains elusive as devices are significantly smaller than the diffraction limit at these wavelengths, hindering far-field spectroscopic tools. Here, we report on the sub-wavelength cavity electrodynamics of graphene embedded in a vdW heterostructure plasmonic microcavity. Using on-chip THz spectroscopy, we observed spectral weight transfer and an avoided crossing between the graphite cavity and graphene plasmon modes as the graphene carrier density was tuned, revealing their ultrastrong coupling. Our findings show that intrinsic cavity modes of metallic gates can sense and manipulate the low-energy electrodynamics of vdW heterostructures. This opens a pathway for deeper understanding of emergent phases in these materials and new functionality through cavity control.

}
%


\maketitle

\section{Main}\label{main-1} 

A vast array of low-energy quantum phenomena have been observed in vdW heterostructures, constructed by mechanically stacking and patterning exfoliated two dimensional (2D) materials. Examples include superconductivity \cite{andrei2020graphene,balents2020superconductivity}, correlated insulating states \cite{mak2022semiconductor}, polaritonic phenomena \cite{koppens2011graphene,grigorenko2012graphene,basov2020polariton,low2017polaritons,du2023moire}, 2D magnetism \cite{huang2020emergent} and both integer and fractional quantum anomalous Hall effects \cite{li2021quantum,park2023observation,zeng2023thermodynamic,cai2023signatures}. These effects are highly sensitive to their electromagnetic environment \cite{song2018electron} and can be tuned \textit{in situ} using electrostatic gates, which control the material's carrier density and electronic structure. We show here that these gates, typically patterned from graphite flakes for ease and improved device quality \cite{zibrov2017tunable}, play an additional role in vdW heterostructures. Due to their sub-wavelength size, they naturally form plasmonic cavities. The edges of the sample set up standing waves of current density, confining light deep in the near-field \cite{graef2018ultra,yoshioka2023chip,zhao2023observation,fei2015edge,lee2019graphene}. The plasmonic cavity resonances of prototypical graphite gates (\SI{\sim 25}{GHz}~-~\SI{2.5}{THz}) coincidentally fall on the same energy scale of many phenomena in vdW heterostructures ($\sim$0.1~-~\SI{10}{\milli\eVolt}). This raises the question: can discrete modes of light confined in a graphite gate modify or even control the electrodynamics of a vdW heterostructure? 

Engineering the properties of quantum materials using the enhanced light-matter interaction provided by a cavity has gained traction as a route to new functionality \cite{
schlawin2022cavity,ruggenthaler2018quantum,hubener2021engineering,eckhardt2023theory,sentef2018cavity,curtis2019cavity,curtis2023local,masuki2023cavity,bloch2022strongly, allocca2019cavity,raines2020cavity,vinas2023controlling, wang2019cavity,masuki2023berry,chakraborty2021long,andolina2024amperean}. When the cavity coupling strength $g$ approaches the cavity resonance frequency $\nu_0$, collective matter modes, such as phonons, plasmons or magnons, are transformed into light-matter hybrids, whose formation can significantly impact the macroscopic properties of the material itself \cite{schlawin2022cavity,hubener2021engineering}. As the interaction becomes stronger and begins to reach the so-called ultrastrong coupling regime, fewer photons, or even only photon vacuum fluctuations, are required to create new thermodynamic ground states \cite{forn2019ultrastrong,frisk2019ultrastrong}. Pioneering experiments have demonstrated an enhancement of ferromagnetism \cite{thomas2021large}, a change in the critical temperature of a metal to insulator transition \cite{jarc2023cavity} and the modification of quantum Hall states in an empty cavity \cite{appugliese2022breakdown}.

 \begin{figure}[b]
    		\centering
      \newcommand*{\figuretitle}[1]{%
    {\centering
    \textbf{#1}
    \par\medskip}
}
      \figuretitle{Fig. 1: Plasmonic microcavity setup for ultrastrong coupling of collective modes in 2D materials}
      \vspace*{+1mm}
    		\includegraphics[]{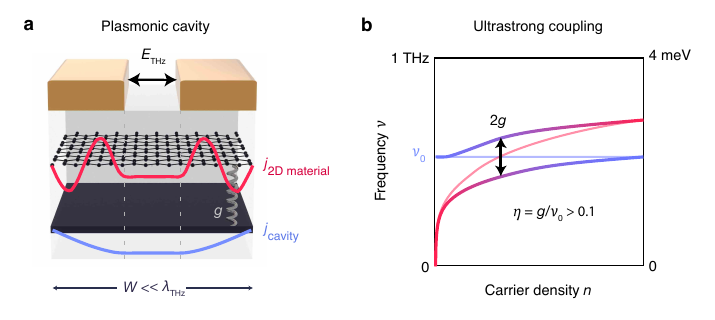}
      \vspace*{+4mm}
    		\caption{\textbf{a} Schematic of a 2D material embedded in a sub-wavelength ($W\ll \lambda_{\text{THz}}$) plasmonic cavity formed by a few-nm graphite flake and surrounding dielectric environment. A collective mode of a 2D material (red), such as a graphene plasmon studied in this work, can hybridize with the plasmonic bare cavity mode (blue). \textbf{b} The frequency of the screened plasmonic mode in the 2D material (light magenta line) is tuned with carrier density into resonance with the screened bare cavity mode, $\nu_0$ (light blue line). The cavity and plasmon modes hybridize due to coupling mediated by unscreened currents between the metal traces and an avoided crossing appears \cite{SupInf}, with an energy splitting of twice the coupling strength, $g$. When $\eta$, the ratio of $g/\nu_0$, exceeds the value of 0.1, the system is in the ultrastrong interaction regime. }

		    \label{fig:figure1}
    \end{figure}

The diverse phenomena found in vdW heterostructures should be particularly sensitive to the plasmonic cavity modes \cite{ashida2023cavity,herzig2024high,dirnberger2023magneto} naturally provided by the metallic gates. Fluctuations are increased due to their reduced dimensionality, preempting the formation of long-range order \cite{auerbach2012interacting,novoselov20162d}. Additionally, the low energy scale of emergent properties makes it easier to attain ultrastrong light matter interaction, where qualitatively new properties could be induced purely by coupling to vacuum fluctuations. However, developing routes for control requires demonstrating and understanding the interaction between cavity modes and a micron-sized vdW heterostructure. Therefore, spectral probes on the meV energy scale are needed, which are experimentally challenging to generate \cite{feurer2007terahertz}. Furthermore, typical device sizes are orders of magnitude smaller than the wavelength of THz light (\SI{\sim 300}{\micro\meter}), precluding far-field measurements.  

Here, we report on the cavity electrodynamics of monolayer graphene embedded in a vdW heterostructure plasmonic microcavity, probed using on-chip THz spectroscopy. We found that the plasmonic collective oscillations in graphene hybridized with the plasmonic cavity mode, native to a microstructured graphite electrostatic gate (see Fig.~\ref{fig:figure1}). We observed spectral weight transfer from the graphite cavity mode to the graphene modes and quantified the normalized coupling strength, $\eta=g/\nu_0$, accessing the evasive ultrastrong light-matter interaction regime ($\eta>0.1$), where cavity engineering becomes possible. We identify the coupling mechanism and provide generalizable design principles for enhancing or minimizing this hybridization, delivering a platform for either engineering low-energy dynamics or spectroscopically probing the undisturbed many-body ground states of quantum materials.

Our findings reveal the general relevance of cavities in vdW heterostructures for tailoring macroscopic properties. This suggests the necessity for understanding the role that the built-in cavity modes play in the known ground state properties of these materials, which could then be applied to develop deterministic cavity control of electronic phases. Such a route heralds the possibility of collective quantum phenomena and new functionality, such as Bose-Einstein condensation of plasmons \cite{berman2010bose,hakala2018bose}, single photon detection in the THz \cite{di2024ultra} or plasmonic quantum networks \cite{bogdanov2019overcoming}. By delivering a cavity platform on a chip that enables deterministic tuning of light-matter interactions, spectral read-out of coupling strength, and simultaneous DC transport to capture the electronic properties of an undriven cavity, this work provides a generalizable, experimental route for investigation and cavity control of emergent phases in vdW heterostructures.

\section{Cavity design and spectral readout}\label{cavity-2} 

\begin{figure}
    		\centering
      \newcommand*{\figuretitle}[1]{%
    {\centering
    \textbf{#1}
    \par\medskip}
}
      \figuretitle{Fig. 2: On-chip Thz spectroscopy of plasmonic microcavity modes}
      \vspace*{+1mm}
    		\includegraphics[]{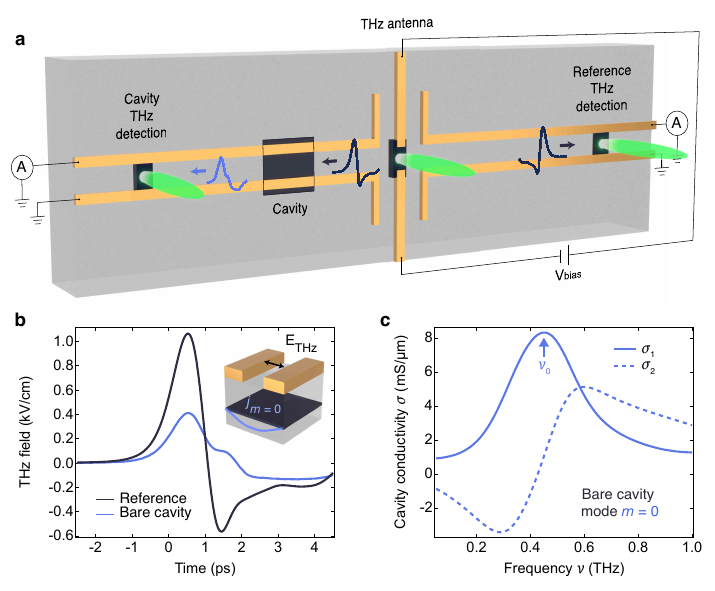}
      \vspace*{+4mm}
    		\caption{\textbf{a} Circuit design for on-chip THz spectroscopy of vdW heterostructure microcavities: Two identical THz pulses (dark blue) are emitted from a voltage-biased THz antenna and coupled into two symmetric coplanar strip transmission line arms. The THz pulse interacts with the graphite cavity (dark $\rightarrow$ light blue) and is measured using a photoconductive switch triggered by a delayed probe beam on the left. The signal can be compared to the reference signal measured on the detector on the right. \textbf{b} Experimentally-measured THz transients of a typical bare cavity (without an additional 2D material) and reference as a function of time delay. For geometric parameters, see \cite{SupInf}. The detected cavity signal is significantly modified as compared to the reference trace. \textbf{c} Real and imaginary parts of the near-field cavity conductivity obtained from the data in b.}
		    \label{fig:figure2new}
    \end{figure}

The cavities reported here combine gate-tunable vdW heterostructure devices of typical dimensions ($\sim10\times$\SI{10}{\micro\meter\squared}) with ultrafast optoelectronic circuitry \cite{mciver2020light}. This circuitry confines light to the near-field, where $(\nu \lambda_{\text{THz}})/ c\ll1$ ($c$ is the speed of light), and enables spectral-readout from 0.1~-~\SI{1}{\THz} \cite{potts2023chip, gallagher2019quantum,zhao2023observation,yoshioka2022ultrafast,yoshioka2023chip,wang2023superconducting,mciver2020light,parker2020chip,zhong2008terahertz,karadi1994dynamic}. Each cavity was constructed using a precision-cut graphite flake, encapsulated in hBN and placed on a sapphire substrate. The hBN preserves the intrinsic properties, like high mobility, of the 2D material from the deleterious effect of the environment \cite{dean2010boron}, and electrically isolates it from the circuitry. These spectrometers feature \textit{in situ} referencing circuitry, allowing the measurement of the time-domain cavity and reference THz pulses on separate transmission line arms (Fig.~\ref{fig:figure2new}a-b). The Fourier transforms of the recorded time-domain signals were then used to calculate the transmission coefficient, which in turn was used for computing the real and imaginary parts of the graphite cavity conductivity, plotted in Fig.~\ref{fig:figure2new}c \cite{SupInf}. 

The experimental bare graphite cavity response in Fig. \ref{fig:figure2new}c exhibits a resonance at $\nu_0$ that is well described by a Lorentz oscillator model and results from the excitation of a 2D plasmonic standing wave, which is predominantly confined beneath the metal traces. The electric field of the THz pulse generates a current in the graphite that reflects at discontinuities in the dielectric environment, particularly at the sample boundaries, forming a standing wave of current density ($j_{m = 0}$ in Fig. \ref{fig:figure2new}b, inset). The wavelength and thus momentum of the confined plasmonic resonance is determined by the geometry and dielectric environment of the heterostructure \cite{graef2018ultra,yoshioka2023chip,zhao2023observation,fei2015edge,lee2019graphene}. We developed an analytical model to describe the plasmonic modes of the bare cavity and their conductivity, whose results match with both experiment and finite element simulations that capture the full electromagnetic field propagation through the cavity \cite{SupInf,svintsov2023refraction}. Thus, the resonance frequency of the cavity mode can be systematically controlled by the microstructured cavity design \cite{SupInf}.

These cavities and associated circuitry provide multiple key functions. First, the circuitry allows for the contact-less, direct measurement of the complex cavity conductivity on the length-scales of typical vdW heterostructure dimensions and spanning the ``THz gap", impossible to access with far-field techniques \cite{feurer2007terahertz}. This allows the extraction of parameters such as linewidths, mobilities and cavity quality factors of collective modes. Importantly, this technique enables both the real and imaginary part of the conductivity to be extracted \cite{nuss2007terahertz}. This is vital for accurate measurements of parameters such as the superfluid stiffness, particularly in systems where electronic correlations shift spectral weight to high frequencies. In such cases, the applicability of using Kramers-Kronig transformations of the absorption to extract complex quantities would be limited \cite{armitage2009electrodynamics,basov2011electrodynamics}. The spectroscopically detected electrodynamics of a vdW heterostructure can additionally be compared to DC transport measurements on the same device \cite{SupInf}. Furthermore, the metallic transmission line strips provide sharp electrodynamic boundary conditions, breaking momentum conservation in a controllable manner that mediates the coupling between the graphite cavity and collective modes in 2D materials, as elaborated below.

\section{Spectrally resolving gate-tunable graphene plasmonic modes}\label{sample-3}

\begin{figure}
    		\centering
         \newcommand*{\figuretitle}[1]{%
    {\centering
    \textbf{#1}
    \par\medskip}
}
      \figuretitle{Figure 3: Probing the spectra of gate-tunable graphene plasmonic modes}
      \vspace*{+1mm}
    		\includegraphics[width=0.62\textwidth]{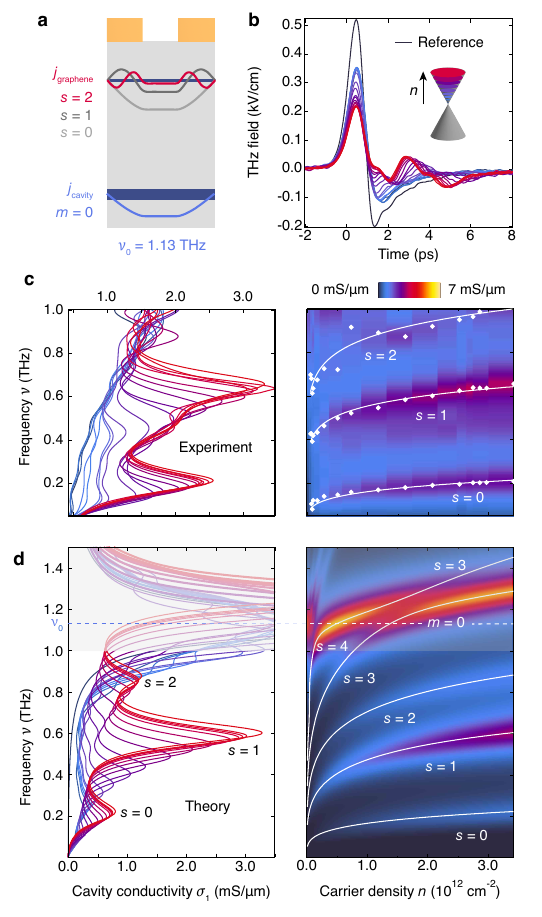}
      \vspace*{+4mm}
    		\caption{\textbf{a} Cavity design for characterizing the plasmonic modes of a graphene layer ($s = 0,1,2$) with a resonance frequency of the bare graphite cavity ($m = 0$) mode, $\nu_0$, of \SI{\sim 1.13}{THz}, outside the experimental spectral sensitivity range (for cavity geometry details see \cite{SupInf}). \textbf{b} Experimentally measured THz transients of the reference (black line) and increasing graphene carrier densities (colored lines). \textbf{c} Experimentally obtained real part of the cavity conductivity spectra. The bare graphite cavity conductivity (dark blue trace) increases slowly in conductivity with increasing frequency, and is expected to peak at $\sim$1.13~THz. The resonance frequencies as a function of graphene carrier density of these three resonances are shown in the right panel. \textbf{d} Analytical model capturing the cavity conductivity for the device shown in (a), identifying the three experimentally detected resonances as the $s = 0,1,2$ graphene modes. The extended frequency range of the theory captures the avoided crossings between the $s = 3,4$ modes to the bare cavity mode (dashed line).}
		   \label{fig:figure3}
    \end{figure}

We investigated the behavior of the plasmonic collective modes of hBN-encapsulated graphene with an electrostatic graphite gate. We designed the device such that the graphite cavity resonance frequency was intentionally centered at \SI{\sim 1.13}{THz}, outside of the experimental spectral sensitivity range, to facilitate the spectral characterization of individual graphene plasmonic modes (Fig.~\ref{fig:figure3}a). On-chip THz data taken at \SI{20}{K} show that the transmission decreases as the sample reflectivity increases with increasing carrier density (Fig.~\ref{fig:figure3}b). Additionally, time-domain oscillations appear, indicating a non-trivial frequency-dependence of the optoelectronic response.

The real (dissipative) part of the cavity conductivity of gate-tunable monolayer graphene embedded in the microcavity was calculated from the time-domain data in Fig.~\ref{fig:figure3}b \cite{SupInf}. The results are plotted in Fig.~\ref{fig:figure3}c. At the lowest carrier density, close to charge neutrality, the cavity conductivity data grow slowly with increasing frequency. This is consistent with the expectation that charge neutral graphene is largely transparent to THz fields. Only the rising edge of the bare cavity resonance is captured, which peaks outside the experimentally accessible spectral range. 

As the graphene becomes more conductive, three resonances appear on top of this background, which grow in amplitude and blueshift as the carrier density increases. This can be explained as the formation of confined plasmonic standing waves in the graphene ($j_{\text{graphene}}$) \cite{graef2018ultra,yoshioka2023chip,zhao2023observation,fei2015edge,lee2019graphene,alonso2017acoustic,xu2023electronic}. Higher frequency resonances, represented schematically in Fig.~\ref{fig:figure3}a, originate from higher order plasmonic modes. The labelling $s = 0, 1, 2$ correspond to the number of nodes in the current density beneath each coplanar strip, as determined from finite element electromagnetic field simulations \cite{SupInf}. The spectra in Fig.~\ref{fig:figure3}c were fit with a sum of Lorentzians to extract the linewidth and resonance frequency of each mode as a function of carrier density. A representative linewidth for the $s=2$ mode at $n=\SI{2.85e12}{\per\centi\meter\squared}$ corresponds to an electronic mean free path of \SI{\sim9.1(0.2)}{\micro\meter}, roughly the width of the device \cite{SupInf}. The corresponding quality factor is $Q\sim{5.7 \pm 0.1}$, defined as $Q\sim\nu_{\textit{s}=2}/\gamma_{\textit{s}=2}$, where $\nu_{\textit{s}=2}$ is the resonance frequency and $\gamma_{\textit{s}=2}$ is the linewidth. The scattering rate and quality factor are limited by device dimensions, illustrating that the graphene in the microstructured cavity exhibited ballistic transport, thereby achieving the highest possible quality factor for a device of this geometry in this frequency range.  

In an uncoupled system and in the absence of electronic correlations, the carrier-density dependence of the resonance frequency is determined by the electronic band dispersion of the sample \cite{hwang2009plasmon}. This can be monitored by tracking the fitted resonance frequency as a function of carrier density $n$. We find that the $s = 0$ mode scales as $\nu\propto n^{0.23\pm 0.03}$, in good agreement with the expectation of a Dirac band structure, where 2D plasmons scale as $\nu \propto n^{0.25}$ \cite{hwang2007dielectric,SupInf}. However, the higher order plasmon modes, $s = 1,2$ scale as $0.12\pm 0.01$ and $0.11\pm 0.01$ respectively. 

To understand the suppressed power laws of the $s = 1,2$ modes, we extended our analytical model to describe the light-matter interaction of a cavity containing a 2D material \cite{SupInf}. Fig.~\ref{fig:figure3}d plots the results of the real part of the simulated cavity conductivity up to \SI{1.5}{THz}, using similar linewidths as extracted in experiment. Three modes appear below \SI{1}{THz} with excellent qualitative agreement to the experimental data. Examining the simulated conductivity outside of the experimental range shows that when the graphene $s = 3,4$ modes cross with the bare cavity mode (dashed line, Fig.~\ref{fig:figure3}d), they hybridize and form avoided crossings, splitting into upper and lower polariton branches. The simulations reveal that at high carrier densities, the observed $s = 1,2$ modes also form lower branches of avoided crossings occurring at carrier densities outside of the experimentally accessible range \cite{SupInf}. The experimentally detected, suppressed power-law behaviors of the graphene $s = 1,2$ modes are thus a consequence of the hybridization of graphene and cavity plasmons.

\section{Ultrastrong cavity-coupling of plasmonic collective modes}\label{USC-4}

\begin{figure}
    		\centering
         \newcommand*{\figuretitle}[1]{%
    {\centering
    \textbf{#1}
    \par\medskip}
}
      \figuretitle{Figure 4: Ultrastrong coupling between plasmonic modes of graphene and cavity}
      \vspace*{+1mm}
    		\includegraphics[width=1\textwidth]{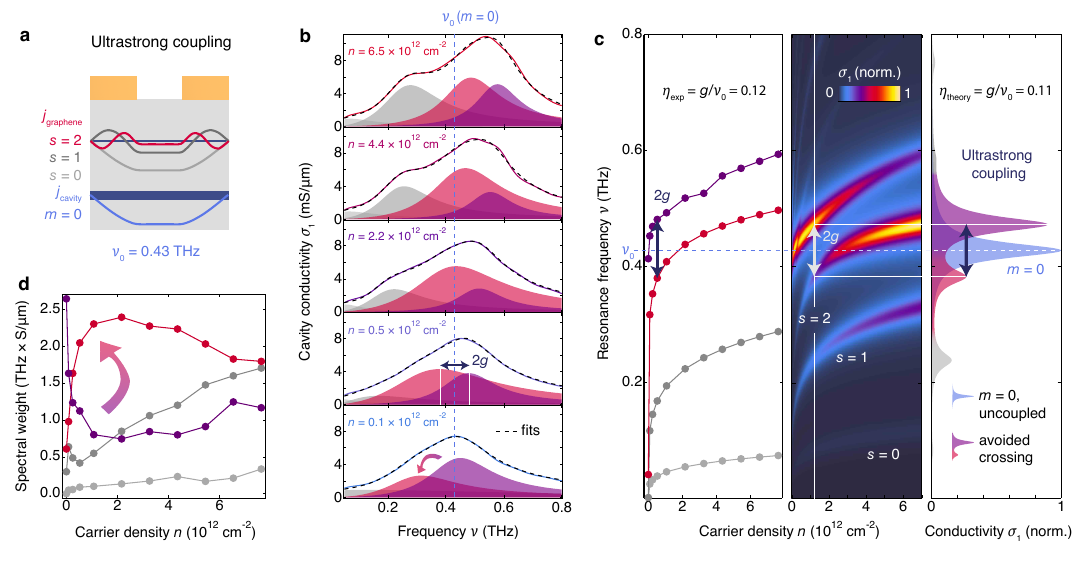}
      \vspace*{+4mm}
    		\caption{\textbf{a} Cavity design with a thin hBN layer between graphene and graphite to center the bare cavity mode to \SI{0.43}{THz} (for cavity geometry details see \cite{SupInf}). The graphene $s = 2$ mode ultrastrongly couples to the cavity $m = 0$ mode inside the experimental spectral sensitivity range. \textbf{b} Experimentally-obtained real part of the near-field THz conductivity spectra of a selection of measured graphene carrier densities. The data were fit to a sum of four Lorentzians as indicated by the shaded peaks. \textbf{c} Left panel: Detected resonance frequencies as a function of carrier density for the four Lorentzians fitted to the experimental data. The experimentally extracted normalized coupling strength $\eta_{\text{exp}} = 0.12\pm 0.01$, achieving the ultrastrong coupling regime. The analytical model for this device (center panel) follows similar trends, with $\eta_{\text{theory}} = 0.11\pm 0.01$. The right panel shows the simulated conductivity spectra at a carrier density of $n = 0$, for the uncoupled $m = 0$ mode and for a carrier density of $n=\SI{1.24e12}{\per\centi\meter\squared}$ at which the avoided crossing appears. \textbf{d} Quantification of experimental spectral weight transfer between the two highest frequency modes.}
		   \label{fig:figure4}
    \end{figure}

Having demonstrated the role of hybridization in a device where resonances were well-separated in frequency, here we report on the observation of ultrastrong coupling between graphene and cavity plasmonic modes within the experimental spectral sensitivity range. To enter this regime, we intentionally designed the cavity such that four modes, the bare cavity mode centered at \SI{0.43}{THz} in addition to three graphene modes, $s = 0,1,2$, overlap in frequency within the experimental measurement range of the THz circuit, achieved by making both the graphite and the hBN between graphene and graphite thinner (Fig.~\ref{fig:figure4}a) as compared to the device in Fig.~\ref{fig:figure3}.

The ultrastrong coupling of graphene plasmonic modes to the cavity mode is evidenced by the near-field carrier-density dependent conductivity spectra, shown in Fig.~\ref{fig:figure4}b. We fit the spectra with a sum of four Lorentzians to capture the four expected modes of the cavity \cite{SupInf}. As the carrier density increases, all modes shift to higher frequencies. At a graphene carrier density of $n = 0.5\times 10^{12}$ cm$^{-2}$, the $s = 2$ mode becomes resonant with the $m = 0$ mode (dashed line in Fig.~\ref{fig:figure4}b). Rather than merely crossing in frequency, as would be expected given the different momenta of the $m = 0$ and $s = 2$ modes \cite{SupInf}, they hybridize and split. The spectral separation of these two modes at this carrier density corresponds to twice the coupling strength $g$, found to be (50~$\pm$~5) GHz, corresponding to \SI{\sim 0.2}{meV}. Comparing this to the frequency of (0.43~$\pm$~0.01) THz, \SI{\sim 1.8}{meV}, at which this coupling occurs, we obtained a normalized coupling strength of $\eta_{\text{exp}} = 0.12 \pm 0.01$ characterizing this cavity as in the ultrastrong coupling regime \cite{SupInf}. 

The ultrastrong coupling between the graphene and cavity plasmonic modes can be seen to significantly modify the electrodynamic properties of the system. Tracking the resonance frequencies of the four modes as a function of carrier density (seen in Fig.~\ref{fig:figure4}c) reveals that the lowest order graphene mode $s = 0$ is again primarily uncoupled from the cavity as it scales with $\nu_{\text{\textit{s}=0}}\propto n^{0.24\pm 0.01}$. However, the $s=1$ and $s=2$ modes are significantly suppressed when compared to their uncoupled nature \cite{SupInf}.

The spectral weight of each mode was calculated as the area of each Lorentzian shown in Fig.~\ref{fig:figure4}b and plotted as a function of carrier density in Fig.~\ref{fig:figure4}d. As the graphene becomes metallic, the spectral weight of the upper branch (purple) decreases, as the lower branch (magenta) rapidly increases in spectral weight. This stands in sharp contrast to the expected spectral weight behavior of graphene and graphite cavity modes in the absence of coupling, where the spectral weight of graphene modes should increase in amplitude and the cavity mode should only be weakly affected when the graphene carrier density $n$ is increased  \cite{SupInf}. The detected transfer in spectral weight across the avoided crossing is thus a consequence of the hybridization and a signature of ultrastrong coupling.

The analytical model captures the carrier-density dependence and spectral weight trends of this device, as shown in the center panel of Fig.~\ref{fig:figure4}c. This theory predicts $\eta_{\text{theory}} = 0.11\pm 0.05$ (right panel), consistent with the experimental observation. The unexpected carrier-density scaling of the plasmon frequency and spectral weight transfer can be attributed to tunable, multimodal couplings between the bare cavity mode and graphene modes, which reach the ultrastrong coupling regime for the $s = 2$ mode \cite{SupInf, tay2023ultrastrong}.

\section{Tunable cavity coupling mechanism}

\begin{figure}
    		\centering
         \newcommand*{\figuretitle}[1]{%
    {\centering
    \textbf{#1}
    \par\medskip}
}
      \figuretitle{Figure 5: Coupling strength tunability through geometry: From cavity sensing to cavity control}
      \vspace*{+1mm}
    		\includegraphics[width=0.7\textwidth]{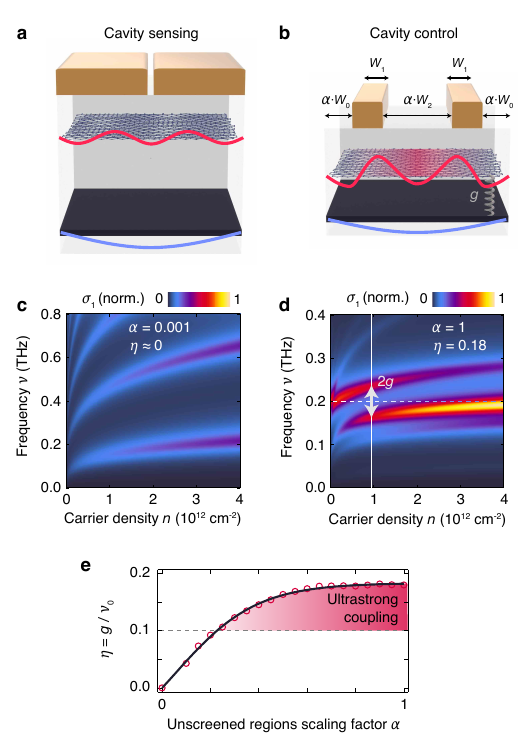}
      \vspace*{+4mm}
    		\caption{\textbf{a} A cavity with a small-gap coplanar strip and no extended sample regions behaves as a sensing tool. \textbf{b} A cavity with extended 2D material region $W_0$ and gap $W_2$ can be used to control collective modes of 2D materials. $\alpha$ represents a tuning factor for the cavity parameters. Coplanar strip geometry parameters of $W_0 = W_1 =$~\SI{3}{\micro\meter} and $W_2 =$~\SI{8}{\micro\meter} were used (for all other device parameters see \cite{SupInf}). \textbf{c} Simulation of a device as in (a) with a graphene monolayer and $\alpha = 0.001$. Multiple undisturbed graphene plasmonic modes can be obtained that all scale as $\nu\propto n^{0.25}$. \textbf{d} Simulation of a device as in (b) with a graphene monolayer and $\alpha=1$. An avoided crossing with $\eta=0.18$ is observed. \textbf{e} $\eta$ dependence on the scaling factor $\alpha$ for the geometry defined in (b), with cavity sensing in the small $\alpha$ limit and cavity control in the large $\alpha$ limit.}

		   \label{fig:figure5}
    \end{figure}

While the devices highlighted in this work exhibited coupling between sample and cavity modes, it would be useful to also be able to build cavities with negligible coupling such that the unperturbed electrodynamics of the sample can be interrogated. The analytical model developed for this work provides design principles to construct cavities which either minimize or maximize coupling strengths to either sense or control the THz electrodynamics of collective modes in vdW materials (Fig.~\ref{fig:figure5}).

The theory predicts that a cavity built with a thick hBN, thick graphite gate, and with a coplanar strip patterned to cover all but a small gap between the traces ($\alpha W_2\ll W_1$, see Fig.~\ref{fig:figure5}a) minimizes the coupling between the graphene and cavity collective modes (Fig.~\ref{fig:figure5}c). In such a device, momentum conservation is preserved and thus the cavity and 2D material plasmon modes are orthogonal to each other. Our theory and numerical simulations show that both the current density $j$ and the spatial derivative, $dj/dx$, are approximately continuous throughout the device \cite{SupInf}. As the THz spectroscopic platform introduced here is capable of independently capturing the real and imaginary parts of the uncoupled 2D material conductivity, which are necessary for extracting spectral weight shifts, measurements of a ``sensing cavity" will provide deeper insight into the dynamics of correlated materials \cite{van2014plasmons}. For example, properties such as quasiparticle scattering rates, eigenfrequencies of collective modes, magnitudes of electronic energy gaps, and the density of superconducting carriers can be captured as a function of temperature, displacement and magnetic field, and carrier density. 

By contrast, the ultrastrong coupling can be enhanced by patterning the dielectric environment (Fig.~\ref{fig:figure5}b). A cavity built with increased conductor separation and additional vdW heterostructure widths outside of the coplanar strip maximizes the ultrastrong coupling between modes of different momenta, with predicted strengths of $\eta > 0.18$ (see Fig.~\ref{fig:figure5}b,d). This is due to broken translational symmetry of this cavity design, and thus broken momentum conservation. While discrete plasmonic and cavity modes are typically orthogonal to each other, the inhomogenous dielectric environment allows graphene plasmonic modes to spatially overlap with the cavity mode, through which they become entangled with each other \cite{tay2023ultrastrong}. Unscreened graphene plasmons are excited in the gap and mediate the coupling of the screened cavity and graphene plasmons. The relative widths of the metal traces to the widths of the unscreened regions provide a tuning knob for the coupling strength. Thus, the coupling can be controlled through the cavity geometry and carrier density, allowing for either probing ($\alpha\ll1$) or controlling ($\alpha\approx1$) quantum materials (see Fig.~\ref{fig:figure5}e). The details of this coupling are captured by our analytical model \cite{SupInf}.

\section{Cavity control of vdW quantum materials}

This work demonstrates the first experimental evidence of the multimodal, ultrastrong interaction between gate-tunable graphene plasmons and a plasmonic graphite cavity in the deep sub-wavelength regime. Our results show that cavity electrodynamics are relevant for vdW heterostructures with graphite gates of typical dimensions. This raises the possibility that cavity effects due to vacuum fluctuations may be relevant in vdW heterostructures exhibiting strongly correlated quantum phases. In common vdW device architectures, microstructured electrostatic gates and contacts may break momentum conservation in a similar way to the metallic coplanar strips employed in this work.

The platform developed here delivers a route to deterministic cavity control of many-body phenomena in vdW heterostructures. Our advances in on-chip THz circuitry allow both the light-matter coupling strength and spectral weight transfer to be probed under the same experimental conditions in which DC transport is measured. With this, the effects of cavity-mediated vacuum fluctuations, as probed by transport, could be experimentally related to the cavity hybridization, as probed by THz measurements. The near-field nature and broken symmetries in these vdW heterostructure microcavities will enable the future investigation of cavity control. This could be performed on not only infrared active, but also optically silent modes \cite{SupInf}, such as acoustic and Raman-active phonons, superconducting collective modes \cite{costa2021harnessing}, and topological edge modes. While the cavity demonstrated here achieves ultrastrong coupling in the linear regime, the possibility to drive the cavity to the nonlinear regime using Floquet engineering protocols may enable additional control in the non-equilibrium domain. With such a strategy, these devices provide a novel and versatile platform for delivering spectroscopic information and transformative cavity-control of emergent properties in vdW heterostructures.

\backmatter
\newpage
\section*{Supplementary information}

\subsection*{Acknowledgements}
We thank A. Cavalleri, D.N. Basov, J.D. Adelinia, E. Wang and M. Chavez-Cervantes for their helpful insight throughout the duration of the project. We also thank J.M. Pizarro, T. Wehling, M. Buzzi, M. Fechner, J. Schmalian and E. Demler for helpful conversations.

\subsection*{Funding}

We acknowledge support from the Max Planck-New York City Center for Non-Equilibrium Quantum Phenomena. G.K. and A.R. acknowledges support by the German Research Foundation through the Cluster of Excellence CUI: Advanced Imaging of Matter (EXC 2056, project ID 390715994). H.M.B., M.H.M. and M.W.D. acknowledge support from the Alexander von Humboldt Foundation. H.M.B. acknowledges financial support from the European Union under the Marie Sklodowska-Curie Grant Agreement no. 101062921 (Twist-TOC). The fabrication of THz circuitry optimization devices was supported by the U.S. Department of Energy, Office of Science, Basic Energy Sciences, under Early Career Award DE-SC0024334. Development of the cavity coupling theory was supported by the Center on Programmable Quantum Materials, an Energy Frontier Research Center funded by the U.S. Department of Energy (DOE), Office of Science, Basic Energy Sciences (BES), under award DE-SC0019443. K.W. and T.T. acknowledge support from the JSPS KAKENHI (Grant Numbers 20H00354 and 23H02052) and World Premier International Research Center Initiative (WPI), MEXT, Japan. A.R. acknowledges support from the Grupos Consolidados (IT1453-22). The Flatiron Institute is a division of the Simons Foundation. At the early stages of the project, G.K. and K.K. were supported by SFB 925 - project 170620586/Deutsche Forschungsgemeinschaft (German Research Foundation). D.M.K acknowledges funding by the Deutsche Forschungsgemeinschaft (DFG, German Research Foundation) within the Priority Program SPP 2244 “2DMP” - 443274199 and under Germany’s Excellence Strategy - Cluster of Excellence Matter and Light for Quantum Computing (ML4Q) EXC 2004/1 - 390534769. M.A.S. was funded by the European Union (ERC, CAVMAT, project no. 101124492).

\subsection*{Duration and volume of study}
This project started in January 2020 and ended in March 2024. During this time, vdW heterostructure microcavity fabrication protocols were developed, the ultrafast cryogenic measurement setup was built, THz circuitry for probing the cavity conductivity was developed and optimized, an analysis framework for extracting the cavity conductivity was constructed, finite element simulations to simulate the cavity modes were performed, and the analytical theory describing the cavity modes and their coupling was developed. For this report, 42 THz devices were measured in total: 15 THz circuitry optimization devices, 16 bare graphite cavities, 3 graphene devices without graphite cavity, 2 cavities based on transition metal dichalcogenide gates and 6 gate-tunable graphene cavity devices.

\subsection*{Author contributions}
J.W.M. conceived the experiment and supervised the project. G.K., B.S., D.H., K.K. and S.K. fabricated the THz cavity devices with help from S.M.L., X.L., J.H., Y.H. and F.S. 
B.S. designed and built the measurement setup. H.M.B. and G.K. performed measurements and analyzed the data, with assistance from M.W.D., K.K. and B.S. H.M.B. developed the data analysis framework. M.H.M. developed the theory with support from A.R. and G.K., and G.K. performed the corresponding numerical simulations. Custom measurement electronics and circuit simulations were provided by T.M. and G.M.. T.T. and K.W. provided high-quality hBN single crystals. E.B., D.M.K., M.A.S., A.R., C.J.E., A.M.P. and G.M. provided valuable conceptual insight. G.K., H.M.B. and J.W.M. wrote the manuscript with contributions from all authors.

\subsection*{Competing interests} 
The authors declare no competing interests.



\end{document}